\documentclass[a4paper,11pt]{article}
\usepackage{jheppub}
\usepackage[utf8]{inputenc}
\usepackage{amsmath}
\usepackage{amssymb}
\usepackage{amsfonts}
\usepackage{mathtools}
\usepackage{amsthm}
\usepackage{graphicx}
\usepackage{slashed}
\usepackage{multirow}
\usepackage{listings}
\usepackage{color}

\usepackage{booktabs}
\usepackage{array,multirow}
\usepackage{dsfont}
\usepackage{url}
\usepackage{subcaption}
\usepackage{changepage}
\usepackage[table]{xcolor}

\usepackage{fancyhdr}

\usepackage[section]{placeins}
\usepackage{float}
\restylefloat{table}

\usepackage{tabularx}
\usepackage{rotating}
\usepackage{etex,etoolbox}
\usepackage{physics}

\usepackage[compat=1.1.0]{tikz-feynman}
\usepackage{arydshln}
\makeatletter
\def\adl@drawiv#1#2#3{%
	\hskip.5\tabcolsep
	\xleaders#3{#2.5\@tempdimb #1{1}#2.5\@tempdimb}%
	#2\z@ plus1fil minus1fil\relax
	\hskip.5\tabcolsep}
\newcommand{\cdashlinelr}[1]{%
	\noalign{\vskip\aboverulesep
		\global\let\@dashdrawstore\adl@draw
		\global\let\adl@draw\adl@drawiv}
	\cdashline{#1}
	\noalign{\global\let\adl@draw\@dashdrawstore
		\vskip\belowrulesep}}
\makeatother
\usepackage{booktabs}
\usepackage{physics}
\usepackage[table]{xcolor}

\usepackage{diagbox}

\usepackage{array,multirow}
\usepackage{bbm}
\usepackage[compat=1.1.0]{tikz-feynman}
\usepackage{minibox}
\usepackage{booktabs, makecell}

\renewcommand{\epsilon}{\varepsilon}
\usepackage{braket}
\usepackage{amsmath}
\usepackage{mathtools}
\usepackage{verbatim}

\usepackage{hyperref}
\usepackage{cleveref}

\usetikzlibrary{decorations.pathreplacing,decorations.markings}

\usepackage{adjustbox}
\usepackage{pifont}
\usepackage{glossaries}
\newacronym{SM}{SM}{Standard Model}
\newacronym{IBP}{IBP}{integration-by-parts}
\newacronym{EW}{weak}{weak}
\newacronym{LO}{LO}{leading-order}
\newacronym{NLO}{NLO}{next-to-leading-order}
\newacronym{PDF}{PDF}{parton distribution function}

\newcommand{\ALOHA}{{\normalfont \scshape\small ALOHA}}

\newcommand{\mgamc}{{\normalfont \scshape\small MG5aMC}}

\newcommand{\UFO}{{\normalfont \scshape\small UFO}}

\newcolumntype{L}[1]{>{\raggedright\let\newline\\\arraybackslash\hspace{0pt}}m{#1}}
\newcolumntype{C}[1]{>{\centering\let\newline\\\arraybackslash\hspace{0pt}}m{#1}}
\newcolumntype{R}[1]{>{\raggedleft\let\newline\\\arraybackslash\hspace{0pt}}m{#1}}
\newcolumntype{N}{@{}m{0pt}@{}}

\numberwithin{equation}{section}

\usepackage{cleveref}

\usepackage[title,titletoc]{appendix}

\definecolor{shaded}{RGB}{245,245,245}
\usepackage{placeins}

\newcommand{\eg}{e.\,g. }
\newcommand{\Eg}{E.\,g. }

\newcommand{\axl}{\mathrm{A}} 
\newcommand{\vcr}{\mathrm{V}} 
\renewcommand{\epsilon}{\varepsilon}
\newcommand\prompt{{\normalfont \ttfamily MG5\_aMC>}} 
\usepackage{listings}
\allowdisplaybreaks[4]

\title{Two-loop amplitude for mixed QCD-EW corrections to $gg \to Hg$}

\author[a]{Matteo Becchetti,}
\author[b]{Francesco Moriello,}
\author[b]{Armin Schweitzer}

\affiliation[a]{Dipartimento di Fisica, Universit\`a di Torino and INFN Sezione di Torino, Via Pietro Giuria 1, I-10125 Torino, Italy}
\emailAdd{matteo.becchetti@unito.it}

\affiliation[b]{ETH Z\"{u}rich, Institut f\"{u}r theoretische Physik, Wolfgang-Pauli
Str. 27, 8093 Z\"{u}rich, Switzerland}
\emailAdd{fmoriell@itp.phys.ethz.ch}
\emailAdd{armin.schweitzer@phys.ethz.ch}

\abstract{We report on the two-loop amplitude computation for the mixed QCD-electroweak corrections to the process $gg \to Hg$, with exact dependence on the electroweak boson masses. This amplitude has been employed in the computation of next-to-leading order (NLO) mixed QCD-electroweak corrections to the Higgs-boson production rate in \cite{Becchetti:2020wof}. The master integrals that appear in the amplitude are evaluated by means of generalized power series  expansions, which allows for fast and high-precision numerical evaluation of the amplitude in the physical phase-space, proving to be a powerful tool for phenomenological applications.}


\begin{document}

\maketitle
\flushbottom



\section{Introduction}
\label{sec:intro}

The discovery of the Higgs boson~\cite{Aad:2012tfa,Chatrchyan:2012ufa} concluded the long ongoing experimental search for the elementary particles described and predicted within the Standard Model (SM) of particle physics.   
This discovery can be seen as the starting point of a precision physics program which aims at the accurate determination of the model parameters and the rigorous assessment of the goodness of the theoretical predictions. \\
Part of this precision program has been focused on studying the Higgs sector, with one important aspect being the
Higgs boson production via gluon fusion at the Large Hadron Collider (LHC) at CERN.
Gluon fusion is by far the dominant Higgs production mode and it is thus of utmost importance to have a very accurate theoretical prediction of this process.\par
The coupling of the Higgs boson to gluons is mediated by a heavy-quark loop.
The Higgs production cross section in gluon fusion was computed at
leading order in the '70s~\cite{Georgi:1977gs},
and at next-to-leading-order (NLO) in the strong coupling constant $\alpha_s$
in the '90s~\cite{Graudenz:1992pv,Spira:1995rr}.
The NLO QCD corrections are sizable ($\sim +100\%$), therefore it is crucial to compute higher-order terms in the perturbative expansion to improve the accuracy of the predictions. \\
The next-to-next-to-leading-order (NNLO)~\cite{Harlander:2002wh,Anastasiou:2002yz,Ravindran:2003um} and the next-next-to-next-to-leading-order (${\rm N^3LO}$)~\cite{
Anastasiou:2013srw,Anastasiou:2013mca,Anastasiou:2014lda,Li:2014afw,
Anastasiou:2015ema,Mistlberger:2018etf} corrections in $\alpha_s$ have been computed in the Higgs Effective Field Theory (HEFT) approach, i.e. in the limit of a top quark much heavier than the Higgs boson, $M_T\gg M_H$. In this limit the loop-mediated coupling is replaced by an effective tree-level one.
The NNLO corrections were found to be significant ($\sim 10 - 20\%$) and with a reduced scale-dependent uncertainty.
The ${\rm N^3LO}$ corrections turn out to be small ($\sim 4 - 6\%$)~\cite{Anastasiou:2016cez}, with a renormalization/factorization scale variation of less than 2\%.\par
Given the very high theoretical accuracy of the ${\rm N^3LO}$ corrections, sub-dominant effects to the Higgs cross section, which are estimated to be in the percent range, have to be considered.\\
One kind of sub-dominant contribution to the cross section is given by the quark-mass effects. 
Firstly, the infinite top mass approximation has a $\sim 6\%$ effect on the SM  cross section with 5 massless flavors and the top. This effect is estimated from the NLO~\cite{Anastasiou:2006hc} prediction, and it can be improved through a multiplicative correction factor applied to the state-of-the-art $\text{N}^3\text{LO}$ HEFT computation.
The finite top-mass contributions mostly factorize from the perturbative corrections~\cite{Anastasiou:2016cez}, so that rescaling results in an estimated $\sim 1 \%$ uncertainty on the prediction~\cite{Pak:2009dg,Harlander:2009my} only. However, this represents a sizeable portion of the remaining theoretical error. At NNLO, top-quark mass effects have been estimated through a power expansion in $M_H/M_T$~\cite{Harlander:2009mq,Pak:2009dg,Harlander:2012hf} to be a $\sim 1\%$ effect. Moreover, very recently \cite{Czakon:2021yub} the NNLO-accurate prediction retaining the full top-mass dependence has been performed, thus effectively removing the residual uncertainty associated with non-factorizing top-mass effects.\\
A different kind of quark-mass effect is given by the contribution stemming from light-quarks. 
At NLO-accuracy, these finite light-quark mass-effects are known exactly~\cite{Spira:1995rr,Harlander:2005rq,Anastasiou:2006hc,Aglietti:2006tp,Bonciani:2007ex,Anastasiou:2009kn,Anastasiou:2020qzk,Anastasiou:2020vkr} and contribute a $\sim -7\%$ change~\cite{Anastasiou:2009kn} to
the cross section, mainly due to top-bottom inteferences. 
Although almost all\footnote{The double-virtual corrections with both quark masses is not known yet.} the relevant ingredients - double-virtual~\cite{Harlander:2019ioe,Czakon:2020vql,Prausa:2020psw}, 
real-virtual~\cite{Bonciani:2016qxi,Bonciani:2019jyb,Frellesvig:2019byn} 
and double-real~\cite{DelDuca:2001fn,Budge:2020oyl} - of the computation including all finite quark-mass effects at NNLO are available, such a computation has not been performed yet, resulting in a remaining residual uncertainty of $\sim 0.8\%$. \\
Beyond the quark-mass effects, another class of suppressed contribution to the Higgs cross section are the so-called ``mixed QCD-electroweak (EW) effects''. They arise at two loops, i.e. at ${\mathcal O}(\alpha^2 \alpha_s^2)$ \footnote{We count all factorized coupling constants except the strong coupling as $\alpha$.}, and are heavily suppressed due to the coupling hierarchy ($\alpha \sim 10^{-1} \alpha_s$). 
They are due to the gluons coupling to EW bosons $V = W, Z$ through a quark loop, followed by the gauge coupling of the EW bosons to the Higgs boson. Mixed QCD-EW contributions were calculated for the light-quark loop~\cite{Aglietti:2004nj,Aglietti:2004ki,Degrassi:2004mx}, for the heavy-quark loop~\cite{Actis:2008ug} and with full quark-mass dependence \cite{Actis:2008ug}, and found to increase the ${\rm N^3LO}$ cross section by about 2\%  \cite{Anastasiou:2016cez}. Since this increase is of the order of the residual QCD uncertainty, it is important to compute the NLO corrections in $\alpha_s$. 
They
consist of three parts: the one-loop $2 \to 3$, the two-loop $2 \to 2$, and  the three-loop $2 \to 1$ with sample diagrams of the last two shown in the first column of \cref{tab:interferences}.
In \cite{Hirschi:2019fkz},
the one-loop $2 \to 3$ processes were computed and found to yield a negligible contribution.
At LO ($\sim {\mathcal O}(\alpha^2 \alpha_s^2)$), the largest part ($\sim 98\%$~\cite{Degrassi:2004mx}) of the mixed QCD-EW contributions
is due to the light-quark part. The evaluation of the NLO ($\sim {\mathcal O}(\alpha^2 \alpha_s^3)$) corrections has, therefore, been aimed at the light-quark part only. 
These corrections were first estimated in the limit where the Higgs mass is much smaller than the EW boson masses, $M_H\ll M_V$~\cite{Anastasiou:2008tj} and they turned out to be sizable. 
The three-loop contribution was evaluated analytically and expressed in terms of multiple polylogarithms (MPLs)~\cite{Bonetti:2016brm}. In \cite{Bonetti:2018ukf}, the soft part of the two-loop $2 \to 2$ process was added, and in \cite{Anastasiou:2018adr} the total cross section was evaluated in the small EW-boson mass limit, $M_V\ll M_H$.
These different approximations gave consistent results. However, they do not allow for a detailed assessment of the remaining uncertainties, since boson-mass and hard effects could not be accessed precisely. Thus, the remaining uncertainty due to mixed QCD-EW contributions to the ${\rm N^3 LO}$-accurate Higgs production cross section in gluon fusion remained a sizeable $\pm 1 \%$ \cite{Anastasiou:2016cez}. This motivates the exact computation of mixed QCD-EW contributions at ${\rm NLO}$. \\
The planar master integrals (MIs) for the two-loop $gg \to Hg$ process with the exact EW-boson mass were published in \cite{Becchetti:2018xsk}
and in \cite{Bonetti:2020hqh} the complete helicity amplitudes, including the non-planar diagrams, were presented. The calculation was done analytically, expressing the results in terms of MPLs. \\
In~\cite{Becchetti:2020wof}, we computed the NLO-accurate corrections to the mixed QCD-electroweak contributions to the Higgs-boson production rate. This computation removed the major uncertainty, i.e. the unknown exact hard-effects at ${\rm NLO}$, and allowed for a significant reduction (almost a factor of two) of the uncertainty associated with mixed QCD-EW contributions to the  gluon fusion Higgs production cross section at ${\rm N^3 LO}$.\par
In this work we augment \cite{Becchetti:2020wof} by providing additional details on the computation of the two-loop amplitude for the partonic process $gg \to Hg$, with the exact EW-boson mass, used in the cross section computation, where we employed the generalized power series expansion method \cite{Francesco:2019yqt} to evaluate the master integrals appearing in the amplitude numerically. This method allows for fast and reliable numerical evaluation of the amplitude in the physical phase-space, proving to be a powerful tool for phenomenological applications \cite{Becchetti:2020wof}. Our result has been checked against ref.~\cite{Bonetti:2020hqh} and we found full agreement. We also provide ancillary material for the numerical evaluation of the MIs, and of the amplitude. While we exploited a private implementation of the generalized power series expansion method, the ancillary material can also be used within the software \textsc{DiffExp} \cite{Hidding:2020ytt} in order to obtain numerical values for the MIs, and thus for the amplitude.\\
The paper is organized as follows. In \cref{sec:results} we summarize the main results of the paper and we describe the general setup of the computation. In \cref{sec:amplitude_computation} we describe in detail the amplitude computation, in particular we discuss the form factors decomposition and consistency checks that have been made to ensure the correctness of the calculation. Finally, in \cref{sec:MIs} we present a brief review of the generalized power series method used to solve the system of differential equations associated to the (MIs) that appear in the amplitude.

\begin{table}[htbp]
\centering
\resizebox{0.9\textwidth}{!}{
\begin{tabular}{c||cc|c}
    \makecell{\diagbox{{\scriptsize EW+QCD}}{{\scriptsize HEFT}}} &
	\makecell{\resizebox{1.5cm}{!}{\begin{tikzpicture}
\begin{feynman}
\vertex(c1);
\vertex[draw,/tikzfeynman/crossed dot,right = 1cm of c1] (c2) {};

\vertex[below right=1.5cm of c2] (b1);
\vertex[above right=1.5cm of c2] (b2);

\diagram* {
	{[edges=gluon]
		(b1) -- (c2) -- (b2),
	},
	{[edges=scalar]
		(c1) -- (c2),
	},
};
\end{feynman}
\end{tikzpicture}}}	
	&\makecell{\resizebox{1.8cm}{!}{\begin{tikzpicture}
\begin{feynman}
\vertex(c1);
\vertex[draw,/tikzfeynman/crossed dot,right = 1cm of c1] (c2) {};

\vertex[below right=1.5cm of c2] (b1);
\vertex[above right=1.5cm of c2] (b2);
\vertex[right=1cm of b1] (d1);
\vertex[right=1cm of b2] (d2);
\diagram* {
	{[edges=gluon]
		(d1) -- (b1) -- (c2) -- (b2) -- (d2),
		(b1) --(b2),
	},
	{[edges=scalar]
		(c1) -- (c2),
	},
};
\end{feynman}
\end{tikzpicture}}}
	&\makecell{\resizebox{1.5cm}{!}{\begin{tikzpicture}
\begin{feynman}
\vertex(c1);
\vertex[draw,/tikzfeynman/crossed dot,right = 1cm of c1] (c2) {};

\vertex[below right=1.5cm of c2] (b1);
\vertex[above right=1.5cm of c2] (b2);
\vertex[above=1 of c1] (d1);
\vertex[above right=.75cm of c2] (d2);
\diagram* {
	{[edges=gluon]
		(b1) -- (c2) -- (b2),
		(d1) -- (d2),
	},
	{[edges=scalar]
		(c1) -- (c2),
	},
};
\end{feynman}
\end{tikzpicture}}}
	\\
	\makecell{\resizebox{2cm}{!}{\begin{tikzpicture}
\begin{feynman}
\vertex (a1);
\vertex[draw,circle,fill=black,inner sep=0pt,minimum size=3pt,right=1cm of a1] (a2) ;
\vertex[draw,circle,fill=black,inner sep=0pt,minimum size=3pt,right=1cm of a2] (a3) ;

\vertex[below=2cm of a1] (b1);
\vertex[draw,circle,fill=black,inner sep=0pt,minimum size=3pt,below=2cm of a2] (b2) ;
\vertex[draw,circle,fill=black,inner sep=0pt,minimum size=3pt,below=2cm of a3] (b3) ;

\vertex[draw,circle,fill=black,inner sep=0pt,minimum size=3pt,below right = 1cm and 1cm of a3] (c1) ;
\vertex[right = 1cm of c1] (c2) {};

\diagram* {
	{[edges=gluon]
		(a1) -- (a2),
		(b1) -- (b2),	
	},
	{[edges=plain]
		(a2) -- (a3) -- (b3) -- (b2) -- (a2),
	},	

		(a3) --[boson,line width=0.5mm] (c1) --[boson,line width=0.5mm] (b3),
	{[edges=scalar]
		(c1) --(c2),
	},
};
\end{feynman}
\end{tikzpicture}}}
	&\cellcolor{red!25} $\mathcal{M}^{(\alpha_s^2\alpha^2)}_{gg\to H}$	& \fbox{\cellcolor{blue!25}$\mathcal{M}^{(\alpha_s^3\alpha^2)}_{gg\to H}$} &	 \\
	\makecell{\resizebox{2cm}{!}{\begin{tikzpicture}
\begin{feynman}
\vertex (a1);
\vertex[right=0.75 of a1] (e1);
\vertex[draw,circle,fill=black,inner sep=0pt,minimum size=3pt,right=1.5cm of a1] (a2) ;
\vertex[draw,circle,fill=black,inner sep=0pt,minimum size=3pt,right=.75cm of a2] (a3) ;

\vertex[below=2cm of a1] (b1);
\vertex[draw,circle,fill=black,inner sep=0pt,minimum size=3pt,below=2cm of a2] (b2) ;
\vertex[draw,circle,fill=black,inner sep=0pt,minimum size=3pt,below=2cm of a3] (b3) ;
\vertex[right=0.75 of b1] (e2);

\vertex[draw,circle,fill=black,inner sep=0pt,minimum size=3pt,below right = 1cm and 1cm of a3] (c1) ;
\vertex[right = 1cm of c1] (c2) {};

\diagram* {
	{[edges=gluon]
		(a1) -- (a2),
		(b1) -- (b2),	
		(e1) -- (e2),
	},
	{[edges=plain]
		(a2) -- (a3) -- (b3) -- (b2) -- (a2),
	},	
		(a3) --[boson,line width=0.5mm] (c1) --[boson,line width=0.5mm] (b3),
	{[edges=scalar]
		(c1) --(c2),
	},
};
\end{feynman}
\end{tikzpicture}}} 			& \fbox{\cellcolor{blue!25}$\mathcal{M}^{(\alpha_s^3\alpha^2)}_{gg\to H}$} 	& $\mathcal{M}^{(\alpha_s^4\alpha^2)}_{gg\to H}$	&  
	\\ \hline
	\makecell{\resizebox{2cm}{!}{\begin{tikzpicture}
\begin{feynman}
\vertex (a1);
\vertex[draw,circle,fill=black,inner sep=0pt,minimum size=3pt,right=1cm of a1] (a2) ;
\vertex[draw,circle,fill=black,inner sep=0pt,minimum size=3pt,right=1cm of a2] (a3) ;
\vertex[right=1.5 of a1] (e1);

\vertex[below=2cm of a1] (b1);
\vertex[draw,circle,fill=black,inner sep=0pt,minimum size=3pt,below=2cm of a2] (b2) ;
\vertex[draw,circle,fill=black,inner sep=0pt,minimum size=3pt,below=2cm of a3] (b3) ;

\vertex[draw,circle,fill=black,inner sep=0pt,minimum size=3pt,below right = 1cm and 1cm of a3] (c1) ;
\vertex[above =1.5 of c1] (e2);
\vertex[right = 1cm of c1] (c2) {};

\diagram* {
	{
		(a1) -- [gluon, red] (a2),
		(b1) -- [gluon, red](b2),
		(e2) -- [gluon, red](e1),
	},
	{
		(a2) --[plain, red] (a3) --[plain, red] (b3) --[plain, red] (b2) --[plain, red] (a2),
	},	
		(a3) -- [boson,line width=0.5mm, red](c1) --[boson,line width=0.5mm, red] (b3),
	{
		(c1) --[scalar, red](c2),
	},
};
\end{feynman}
\end{tikzpicture}}} &
	\multicolumn{2}{c}{} & \fbox{\cellcolor{blue!25}$\mathcal{M}^{(\alpha_s^3\alpha^2)}_{gg\to Hg}$}	 
\end{tabular}
}
\caption{Overview of the relevant interferences necessary for the computation of the cross sections $\sigma^{(\alpha_s^2 \alpha^2 + \alpha_s^3\alpha^2)}_{gg\to H+X}$ presented in \cite{Becchetti:2020wof}. The red colored cell is the LO and cells highlighted in blue are part of the NLO contribution. Amplitudes are denoted by a single representative diagram. Curly lines denote gluons, wavy lines massive weak gauge bosons, continuous straight lines massless quarks and the dashed line represents the Higgs boson. The red diagram at the bottom left is a representative for the two-loop amplitude considered in this work. Diagrams are drawn with TikZ-Feynman \cite{Ellis:2016jkw}. \label{tab:interferences}}
\end{table}

\section{Overview of the computation}
\label{sec:amplitude}

The main result of this paper is the computation of the two-loop amplitude for the partonic process $g(p_1) g(p_2) \to g(p_3)  H(p_4)$. This is a necessary ingredient for the light-quark contribution to the NLO mixed QCD-electroweak (EW) corrections to Higgs production in gluon fusion at the LHC with exact dependence on the EW gauge boson masses \cite{Becchetti:2020wof}. \\

The amplitude for the partonic process $g(p_1) g(p_2) \to g(p_3) H(p_4)$ can be written as
\begin{align} \label{eq:amp}
    \mathcal{A} &= 
    - \frac{i}{2} f_{a_1 a_2 a_3} 
    \left( - \frac{ \sqrt{\alpha_{EW}^3 \alpha_s^3} }{4 \pi s_w^3 } m_W \right)
    \epsilon_{\mu_1}(p_1,p_2)\epsilon_{\mu_2}(p_2,p_1) \epsilon_{\mu_3}^{*} (p_3,p_1) \nonumber \\
    & \times \sum \limits_{i=1}^{4} T_i^{\mu_1 \mu_2 \mu_3}
    \left(\sum \limits_{V=W,Z} \frac{\kappa_V}{m_V^4} A_i\left(\frac{s}{m_V^2},\frac{t}{m_V^2},\frac{m_H^2}{m_V^2}\right) \right)
\end{align}
with $p_1^2=p_2^2=p_3^2=0$, $p_4^2=m_H^2$, $s=(p_1+p_2)^2$, $t=(p_1-p_3)^2$, $m_H$ ($m_V$) is the Higgs (weak) boson mass, $\epsilon(p_k,q_k)$ are polarization vectors for the external gluon $k$ with reference momentum $q_k$, and the $T_i$ are gauge-invariant Lorentz tensors derived in \cref{subsec:tens_decomp}. The global couplings due to light-quark contributions (see \cref{subsec:coupling_struc}) are
\begin{align}
    \kappa_W = 1, &&
    \kappa_Z = \frac{1}{c_W^4} \left( \frac{11 s_W^4}{9}-\frac{7 s_W^2}{6}+\frac{5}{8} \right),
\end{align}
where $\alpha_s$ ($\alpha_{EW}$) denotes the strong (weak) coupling constant and $s_W$ ($c_W$) the sine (cosine) of the Weinberg angle. The $A_i\left(\frac{s}{m_V^2},\frac{t}{m_V^2},\frac{m_H^2}{m_V^2}\right)$ are functions of the rescaled kinematic invariants and of the master integrals (MIs) defined in \cref{sec:amplitude_computation}, which we provide in the ancillary material.\par
The other main result for this paper is the evaluation of the relevant MIs for the two-loop amplitude \eqref{eq:amp} by means of the generalized power series expansion method \cite{Francesco:2019yqt}. The starting data for this method are the knowledge of the system of differential equations and a set of boundary points for the MIs; then the generalized series technique allows us to transport the boundary values to a new kinematic point. This method allows for fast high-precision numerical evaluation which can be improved by having a precomputed grid of boundary points.
Moreover, the analytic continuation of the generalized series expansion in the physical region is completely algorithmic.
\label{sec:results}
\section{Computation of the amplitude}
\label{sec:amplitude_computation}
In this section we describe the details of the amplitude computation for the process under consideration. To obtain the amplitude we generate all relevant Feynman diagrams with QGraf \cite{Nogueira:1991ex}, perform the color and Lorentz algebra with private computer codes and decompose it into a set of four gauge-invariant form factors. We then map all diagrams to a minimal set of two propagator structures and perform an Integration-by-Parts (IBP) reduction \cite{Chetyrkin:1981qh} to a minimal set of master integrals (MIs) with the computer code Kira~\cite{Maierhoefer:2017hyi,Maierhofer:2018gpa}. The MIs are computed as described in \cref{sec:MIs}. All computations are performed in conventional dimensional regularization (CDR) \cite{collins_1984}. \\
In the first \cref{subsec:coupling_struc} we describe the decomposition of the amplitude into four form factors based on the coupling structure of the process, and we show that actually just one form factor, $\mathcal{A}_{\vcr\vcr}$, has to be explicitly computed. In \cref{subsec:tens_decomp} we perform the tensor decomposition of $\mathcal{A}_{\vcr\vcr}$. Finally, in the third \cref{subsec:func_rel}, we discuss the derivation of functional relations exploited to simplify the amplitude, and in the last one, \cref{subsec:AVV_checks}, the checks performed to validate our result.

\subsection{Coupling structure of the light-quark contribution}
\label{subsec:coupling_struc}
In the following we outline the coupling structure of the light-quark contribution \cref{eq:amp}. 
In order to discuss the decomposition we separate the couplings of
the quarks to the electroweak gauge bosons in SM as follows:
\begin{align}
      u d W^+&\propto g_{\vcr,W}\gamma^\mu + g_{\axl,W} \gamma^\mu \gamma^5\,, \nonumber \\
      u d W^-&\propto g_{\vcr,W}^* \gamma^\mu + g_{\axl,W}^* \gamma^\mu \gamma^5\,,      \\
      q q Z&\propto g_{\vcr,Z}  \gamma^\mu+g_{\axl,Z} \gamma^\mu \gamma^5\,. \nonumber
\end{align}
We refer to $g_\vcr$ as the vector coupling constant, and to $g_\axl$ as the axial coupling
constant. Following this coupling separation we may write the mixed
QCD-EW amplitude as:
\begin{align} \label{eq:A}
    \mathcal{A}
     & =\mathcal{A}_{\vcr\vcr} + \mathcal{A}_{\axl\vcr} + \mathcal{A}_{\vcr\axl} + \mathcal{A}_{\axl\axl} \nonumber \\
     & =
    \sum \limits_{vb = (ZZ, W^\pm W^\mp)} g_{H,vb} \left(g_{V,vb}^2 A_{\vcr\vcr} +g_{A,vb}g_{V,vb} A_{\axl\vcr}
    + g_{V,vb}g_{A,vb} A_{\vcr\axl}+g_{A,vb}^2 A_{\axl\axl}\right),
\end{align}
where $g_{H,vb}$ is the coupling of the weak bosons to the Higgs. 
This decomposition highlights the coupling structure of the electroweak loop with representative diagrams shown in the first column of \cref{tab:interferences}.
In order to compute the  mixed QCD-EW cross section, we are interested in the interference of electroweak amplitudes against the pure QCD background shown in the first row of \cref{tab:interferences}. Therefore, for the cross section under consideration, the mixed coupling structures $A_{\vcr\axl}$ and $A_{\axl \vcr}$
are of no concern, since they will not contribute to the interferences.
In particular, the relevant amplitude can thus be written as:
\begin{align} \label{eq:A_relevant}
    \mathcal{A}
     & =\mathcal{A}_{\vcr\vcr} + \mathcal{A}_{\axl\axl}  =
    \sum \limits_{vb = (ZZ, W^\pm W^\mp)} g_{H,vb} \left(g_{V,vb}^2 A_{\vcr\vcr} +g_{A,vb}^2 A_{\axl\axl}\right).
\end{align}
This also implies that the subtleties arising in the embedding of $\gamma^5$ into CDR do not arise in our computation, since the $\gamma^5$-odd traces do not contribute\footnote{They appear in the neglected $A_{\vcr \axl (\axl\vcr)}$-pieces only}. We can treat $\gamma^5$ as completely anti-commuting and one can show that the
pure vector-piece $A_{\vcr\vcr}$ and the pure axial-piece $A_{\axl\axl}$ are equal:
\begin{align}
    A_{\vcr \vcr} =  A_{\axl \axl} .
\end{align}
This is due to the fact that all the relevant $\gamma$-chains of $A_{\axl\axl}$ are of the form:
\begin{align}
 \gamma^5 \gamma^{\mu_1} \dots \gamma^{\mu_{2n}} \gamma^5 = \gamma^{\mu_1} \dots \gamma^{\mu_{2n}}, 
\end{align}
where $n$ is an integer.\par
The second simplification of the amplitude computation arises solely from phenomenological considerations. As already alluded to in the introduction, the top-quark contribution to the mixed QCD-EW cross section at LO makes up only $\sim 2\%$ of the contribution. It is reasonable to expect a similar behaviour at NLO and we therefore restrict ourselves to the computation of the light-quark contributions, \eg 5 massless flavors. Removing the top-quark will manifestly break gauge invariance, since the
$SU(2)$-doublet involving the left-handed top is effectively removed from the computation. We implement it in practical terms by restricting the $W$-exchange contribution to a diagonal
mixing matrix where we neglect the top-bottom flavour exchange, such that:
\begin{align}
    {g_{\axl, W^\pm W^\mp}^2}={g_{\vcr, W^\pm W^\mp}^2}=-\frac{1}{4}\frac{g^2}{2}\sum_{\{(u,d),(c,s)\}} 1 = - \frac{1}{4} g^2 \ ,
\end{align}
while for the $Z$-boson exchange we include the bottom quark and obtain
\begin{align}
    {g_{\axl,ZZ}^2} & =-\frac{1}{4}\frac{g^2}{c_W^2} \sum_{f=\{u,c,d,s,b\}}\left( T^3_{f}\right)^2 = -\frac{5 }{16 c_W} g^2 \,,
    \nonumber                                                                                                      \\
    {g_{\vcr,ZZ}^2} & =-\frac{1}{4}\frac{g^2}{c_W^2} \sum_{f=\{u,c,d,s,b\}}\left( T^3_{f}-2 Q_f s_w^2\right)^2 = -\frac{\left(176 s_w^4-168 s_w^2+45\right)}{144 c_W^2} g^2 \, ,
\end{align}
where $s_W$ ($c_W$) denotes the sine (cosine) of the Weinberg mixing angle, $T^3_{f}$ the weak
isospin, and $Q_f$ the electric charge.\par 
Following the previous discussion, when we compute mixed
amplitudes, we are concerned with the computation of the pure vector piece
$A_{\vcr\vcr}$ for an arbitrary vector boson with mass $m_V$ and a massless
quark.
The $A_{\axl\axl}$ pieces, the relevant quark flavours, and the $W$- and $Z$-bosons are then restored
by inserting the associated couplings in \cref{eq:A_relevant}:
\begin{align}
\mathcal{A}
     &  =
    \sum \limits_{vb = (ZZ, W^\pm W^\mp)} g_{H,vb} \left(g_{V,vb}^2 A_{\vcr\vcr} +g_{A,vb}^2 A_{\axl\axl}\right) \nonumber \\
  & = g_{H,ZZ} \left(g_{V,ZZ}^2 + g^2_{A,ZZ}\right) A_{VV}(m_Z) + 2 g_{H,WW} \left(g_{V,WW}^2 + g^2_{A,WW}\right) A_{VV}(m_W) \nonumber \\
   &=- g^3 m_W \left(  \frac{1}{c_W^4} \left(\frac{11 s_w^4}{9}-\frac{7 s_w^2}{6}+\frac{5}{8}\right) A_{VV}(m_Z)
    + A_{VV}(m_W) \right),
\end{align}
where the factor two accounts for the $W^+ W^-$ and $W^- W^+$ configuration and $\alpha_{EW} = \frac{g^2 s_w^2}{4 \pi}$. 
Including the color-factors we arrive at the coupling structure in \cref{eq:amp}.

\subsection{Gauge invariant tensor decomposition }
\label{subsec:tens_decomp}
In order to perform the analytic computation one is interested in decomposing the amplitude into a minimal set of gauge-invariant tensor structures defined by the external particles of the process under consideration. Such a tensor decomposition involves the analytic solution of potentially large systems. Here, the large size of these systems is mainly due to the regularization scheme choice. When we work in CDR, we lift all structures to $d$-dimensions. Such a treatment has many advantages, \eg renormalization constants are particularly easy, and one does not have to treat external structures differently from off-shell structures. However, it also comes with drawbacks, \eg one can not easily define explicit helicity states, and the lift of the Dirac-structures becomes non-trivial since one does not have a finite basis as for the four-dimensional case. In comparison, for example, the 't~Hooft-Veltman scheme keeps the external structure, \eg external momenta, polarizations, and spinors, in strictly four dimensions. It explicitly splits the algebra and the loop-momenta into 4-dimensional, and $-2\epsilon$-dimensional orthogonal components, and requires introducing additional renormalization pieces, which account for this splitting. However, keeping the external states strictly four-dimensional has advantages. In particular, one can work with physical, on-shell amplitudes, \eg defined helicity states, which gets rid of unphysical, spurious structures. Such an approach can simplify the construction of the projectors by not projecting on generic, Ward-identity fulfilling Lorentz structures, but specific helicity states. This was put forward in a general approach recently in \cite{Peraro:2019cjj,Peraro:2020sfm} and \cite{Chen:2019wyb}, and we refer to the discussion and references therein for more details. \\
In our computation we do not work within the framework of helicity amplitudes. Instead we follow the more ``traditional'' multi-loop approach, in which a set of projectors for generic helicities is obtained, see \eg \cite{Abreu:2018jgq, Glover:2004si,Gehrmann:2011aa}, that can be applied to project the amplitude onto a minimal set of Ward-identity fulfilling, independent tensor structures. The derivation of the decomposition for the amplitude under consideration is detailed in the following.

The amplitude for the vector piece of the process $g(p_1) \ g(p_2) \to g(p_3) \ H(p_4)$ may be written as:
\begin{align} 
	A_{\vcr \vcr} = \sum_{i=1}^{14} t_i S_{i} \label{eq:avv_first}
\end{align}
where the $t_i$ are all possible rank-three Lorentz-tensors obtained from the metric and the external momenta, and the $S_i$ are scalar loop-integrals independent of the polarization vectors. 
The tensor-structures are: 
\begin{align}
    &p_{i\neq 1}^{\mu_1}p_{j\neq 2}^{\mu_2} p_{k\neq 3}^{\mu_3}|_{i \neq j \neq k \neq i} 
    :
    &&t_1 = (\epsilon_1 p_2) (\epsilon_2 p_3) (\epsilon_3^* p_1) 
    && t_2 = (\epsilon_1 p_3) (\epsilon_2 p_1) (\epsilon_3^* p_2)  
     \\ 
    &p_{i\neq 1}^{\mu_1}p_{1}^{\mu_2} p_{1}^{\mu_3}
    :
    &&t_3 = (\epsilon_1 p_2) (\epsilon_2 p_1) (\epsilon_3^* p_1) 
    && t_4 = (\epsilon_1 p_3) (\epsilon_2 p_1) (\epsilon_3^* p_1)  
     \\ 
    &p_{2}^{\mu_1}p_{i \neq 2}^{\mu_2} p_{2}^{\mu_3}
    :
    &&t_5 = (\epsilon_1 p_2) (\epsilon_2 p_1) (\epsilon_3^* p_2) 
    && t_6 = (\epsilon_1 p_2) (\epsilon_2 p_3) (\epsilon_3^* p_2) 
     \\ 
    &p_{3}^{\mu_1}p_{3}^{\mu_2} p_{i\neq 3}^{\mu_3}
    :
    &&t_7 = (\epsilon_1 p_3) (\epsilon_2 p_3) (\epsilon_3^* p_1) 
    &&t_8 = (\epsilon_1 p_3) (\epsilon_2 p_3) (\epsilon_3^* p_2)  
    \\
    & g^{\mu_1 \mu_2} p_{i \neq 3}^{\mu_3} :
    &&t_{9} = (\epsilon_1 \epsilon_2) (\epsilon_3^* p_1) 
    && t_{10} = (\epsilon_1 \epsilon_2) (\epsilon_3^* p_2) 
     \\
    & g^{\mu_1 \mu_3} p_{i \neq 2}^{\mu_2} :
    && t_{11} =  (\epsilon_1 \epsilon_3^*) (\epsilon_2 p_1) 
    &&t_{12} = (\epsilon_1 \epsilon_3^*) (\epsilon_2 p_3) 
     \\
    & g^{\mu_2 \mu_3} p_{i \neq 1}^{\mu_1} :
    && t_{13} = (\epsilon_2 \epsilon_3^*) (\epsilon_1 p_2)  
    && t_{14} = (\epsilon_2 \epsilon_3^*) (\epsilon_1 p_3) 
\end{align}
where the first element of each line states the Lorentz-structure that gives rise to the $t_i$ and transversality of the polarizations $(p_i \epsilon_i )=0$ is imposed.

Requiring gauge invariance,
\begin{align}
	A_{\vcr \vcr}|_{\epsilon(p_i) \to p_i } =0 \, , \qquad i=1,\dots,3 \,,
\end{align}
one finds relations between the $S_i$, and the amplitude may be written as 
\begin{align}
	A_{\vcr \vcr} &= \sum\limits_{i=1}^{4} T_i A_i \nonumber\\
	&= \sum\limits_{i=1}^{4} \left( \epsilon_{\mu_1}(p_1)\epsilon_{\mu_2}(p_2) \epsilon_{\mu_3}^{*} (p_3) T_i^{\mu_1 \mu_2 \mu_3} \right) A_i  \label{eq:vector_vector_piece} \,,
\end{align}
where the $A_i$ are linear combinations of the $S_i$ in \cref{eq:avv_first}. 
The $T_i$ fulfill Ward identities independently. Their components are
\begin{align}
        T_1^{\mu_1 \mu_2 \mu_3} &=-\frac{\left(s_{23} p_1^{\mu_3}-s_{13} p_2^{\mu_3}\right) \left(s_{12} g^{\mu_1 \mu_2}-2 p_2^{\mu_1} p_1^{\mu_2}\right)}{2 s_{23}} ,\\
        T_2^{\mu_1 \mu_2 \mu_3} &= -\frac{\left(s_{13} p_2^{\mu_1}-s_{12} p_3^{\mu_1}\right) \left(s_{23} g^{\mu_2 \mu_3}-2 p_3^{\mu_2} p_2^{\mu_3}\right)}{2 s_{13}} ,\\
        T_3^{\mu_1 \mu_2 \mu_3} &= -\frac{\left(s_{23} p_1^{\mu_2}-s_{12} p_3^{\mu_2}\right) \left(s_{13} g^{\mu_1 \mu_3}-2 p_3^{\mu_1} p_1^{\mu_3}\right)}{2 s_{23}} , \\
        T_4^{\mu_1 \mu_2 \mu_3} &= \frac{1}{2} \left(g^{\mu_2 \mu_3}\left(s_{13} p_2^{\mu_1} -s_{12} p_3^{\mu_1}  \right) +g^{\mu_1 \mu_3} \left(s_{12} p_3^{\mu_2}-s_{23} p_1^{\mu_2}\right)
        \right. \nonumber\\
            &\left. \qquad
        +g^{\mu_1 \mu_2} \left(s_{23} p_1^{\mu_3}-s_{13} p_2^{\mu_3}\right) -2\left( p_2^{\mu_1} p_3^{\mu_2} p_1^{\mu_3}- p_3^{\mu_1} p_1^{\mu_2} p_2^{\mu_3}\right)\right)  \label{eq:tensor_structures_ggHg} ,
\end{align}
with $s_{ij}=2 \left(p_i\,p_j\right)$. \\
These tensor-structures are not unique. In order to see this, consider 
$A_{VV}|_{\epsilon_1 \to p_1} = 0$ from \cref{eq:avv_first}:
\begin{align}
    A|_{\epsilon_1 \to p_1} = 0& \\
    \Rightarrow
    0 &=    
    \left[ S_6 \left(p_1 p_2\right)+S_8 \left(p_1 p_3\right)\right] \left(\epsilon_2 p_3\right) \left(\epsilon_3^* p_2\right) 
    \nonumber 
    \\
    &+\left[ S_2 \left(p_1 p_3\right)+S_5 \left(p_1 p_2\right)+S_{10}\right] \left(\epsilon_2 p_1\right) \left(\epsilon_3^* p_2\right) 
    \nonumber \\
    &
    +\left[S_3 \left(p_1 p_2\right)+S_4 \left(p_1 p_3\right)+S_9+S_{11}\right] \left(\epsilon_2 p_1\right) \left(\epsilon_3^* p_1\right) 
    \nonumber 
    \\
    &
    +\left[S_1 \left(p_1 p_2\right)+S_7 \left(p_1 p_3\right)+S_{12}\right] \left(\epsilon_2 p_3\right) \left(\epsilon_3^* p_1\right)
    \nonumber \\
    &+ \left[S_{13} \left(p_1 p_2\right)+S_{14} \left(p_1 p_3\right)\right]\left(\epsilon_2 \epsilon_3^*\right) .
\end{align}
Here, each term in the square-brackets, multiplying the contracted polarization, has to vanish resulting in five relations among linear combinations of the $S_i$. Once we impose the other Ward identities as well, we get a total of 15 equations (5 from each Ward-identity). Some of the relations will be linear combinations of others and in order to solve the overdetermined system efficiently, we introduce a strict ordering $S_{i}\prec S_{j}$ if $i < j$ and solve w.r.t. the variable with the highest ordering. This is in complete analogy to the well known Laporta-algorithm used in the integration-by-parts (IBP) reduction to scalar MIs.  For example, from the first Ward-identity above we get, using this particular ordering:
\begin{align}
    &S_8=-\frac{S_6 \left(p_1 p_2\right)}{\left(p_1 p_3\right)}, &&
    S_{10}= - S_2 \left(p_1 p_3\right)-S_5 \left(p_1 p_2\right) ,\\
    &S_{11}= - S_3 \left(p_1 p_2\right)-S_4 \left(p_1 p_3\right)-S_9, &&
    S_{12}= -S_1 \left(p_1 p_2\right)-S_7 \left(p_1 p_3\right) ,\\
    &S_{14}=-\frac{S_{13} \left(p_1 p_2\right)}{\left(p_1 p_3\right)}.
\end{align}
As for the IBP-reduction, the choice of the ordering will define a different set of independent $S_i$ (similar to the case of MIs), which will ultimately result in different gauge invariant Lorentz-tensors $T_i$. In particular, some ordering choices may give a ``better'' (\eg more compact) definition of the basis of tensor structures than others. 

To extract the scalar components $A_i$ from the amplitude  $A_{\vcr \vcr}$ (see \cref{eq:vector_vector_piece}), one can construct ``projectors'', denoted as $P_i$ such that
\begin{align}
    \langle P_i ,T_j \rangle &:= \sum_{\text{hel.}} \left(\epsilon^{*}_{\nu_1}(p_1)\epsilon^{*}_{\nu_2}(p_2) \epsilon_{\nu_3} (p_3) P_i^{\nu_1 \nu_2 \nu_3}\right) \left( \epsilon_{\mu_1}(p_1)\epsilon_{\mu_2}(p_2) \epsilon_{\mu_3}^{*} (p_3) T_j^{\mu_1 \mu_2 \mu_3}  \right) \nonumber \\
    &= \delta_{ij} \,. 
\end{align}
In particular, if the $T_i$ are a complete set of linearly independent Lorentz structures for the process under consideration, the projectors may be decomposed as
\begin{align}
    P_k = \sum_{i=1}^{4} c_{k,i} T_i ,
\end{align}
where the $c_{k,i}$ are rational functions of the Mandelstam variables and the dimension.
Denoting the reference vector for the external momentum $p_i$ as $q_i$  one can consider
\begin{align}
    \langle P_i ,T_j \rangle 
    &=
    \sum_{\text{hel.}} \left(\epsilon^{*}_{\nu_1}(p_1)\epsilon^{*}_{\nu_2}(p_2) \epsilon_{\nu_3} (p_3) P_i^{\nu_1 \nu_2 \nu_3}\right) \left( \epsilon_{\mu_1}(p_1)\epsilon_{\mu_2}(p_2) \epsilon_{\mu_3}^{*} (p_3) T_j^{\mu_1 \mu_2 \mu_3} \right) \nonumber\\
    &= 
    \sum_{k=1}^{4} c_{i,k} T_k^{\nu_1 \nu_2 \nu_3} 
    \left( -g_{\mu_1 \nu_1} + \frac{(p_{1})_{\mu_1}(q_{1})_{\nu_1}+(q_{1})_{\mu_1}(p_{1})_{\nu_1} }{(p_1 q_1)} \right) 
    \nonumber\\
    &\phantom{=\sum_{k=1}^{4} c_{i,k} T_k^{\nu_1 \nu_2 \nu_3} } 
    \left( -g_{\mu_2 \nu_2} + \frac{(p_{2})_{\mu_2}(q_{2})_{\nu_2}+(q_{2})_{\mu_2}(p_{2})_{\nu_2} }{(p_2 q_2)} \right)
    \nonumber\\
    &\phantom{=\sum_{k=1}^{4} c_{i,k} T_k^{\nu_1 \nu_2 \nu_3} }
    \left( -g_{\mu_3 \nu_3} + \frac{(p_{3})_{\mu_3}(q_{3})_{\nu_3}+(q_{3})_{\mu_3}(p_{3})_{\nu_3} }{(p_3 q_3)} \right)
    T_j^{\mu_1 \mu_2 \mu_3} \nonumber\\
    &= 
    \sum_{k=1}^{4} c_{i,k} T_k^{\nu_1 \nu_2 \nu_3} 
    \left( -g_{\mu_1 \nu_1}\right) 
    \left( -g_{\mu_2 \nu_2}\right) 
    \left( -g_{\mu_3 \nu_3} \right)
    T_j^{\mu_1 \mu_2 \mu_3} \,,
\end{align}
where we used that by construction $T|_{\epsilon^{*}(p_i) \to p_i } =T|_{\epsilon(p_i) \to p_i } =0$.
Thus, $\langle P_i ,T_j \rangle$ is independent of the chosen reference momentum, and it is sufficient that
\begin{align}
	(P_{i,\mu_1 \mu_2 \mu_3} T_j^{\mu_1 \mu_2 \mu_3}) = -\delta_{ij} ,
\end{align}
where the $-1$ is a direct consequence of exchanging polarization sums of the inner product $\langle \cdot, \cdot \rangle$ for Lorentz-contractions of the respective coefficients.
The components of the ``projectors'' $P_i$ can then be constructed as
\begin{align}
	P_i^{\mu_1 \mu_2 \mu_3} =- (B^{-1})_{ij} T_j^{\mu_1 \mu_2 \mu_3} \,,
\end{align}
where 
\begin{align}
	B_{ij} = (T_{i,\mu_1 \mu_2 \mu_3} T_j^{\mu_1 \mu_2 \mu_3}) \,,
\end{align}
since 
\begin{align}
	P_i^{\mu_1 \mu_2 \mu_3}T_{j,\mu_1 \mu_2 \mu_3} =- (B^{-1})_{ik} T_k^{\mu_1 \mu_2 \mu_3} T_{j,\mu_1 \mu_2 \mu_3} = -(B^{-1})_{ik} B_{kj} = - \delta_{ij} .
\end{align}
For the tensor structures in \eqref{eq:tensor_structures_ggHg} one finds, following this procedure, the components of the projectors:
\begin{align}
        -P_1^{\mu_1 \mu_2 \mu_3} &= -\frac{d s_{23} T_1^{\mu_1 \mu_2 \mu_3}}{(d-3) s_{12}^3 s_{13}}-\frac{(d-4) T_2^{\mu_1 \mu_2 \mu_3}}{(d-3) s_{12}^2 s_{23}}+\frac{(d-4) s_{23} T_3^{\mu_1 \mu_2 \mu_3}}{(d-3) s_{12}^2 s_{13}^2} \nonumber \\
        &-\frac{(d-2) T_4^{\mu_1 \mu_2 \mu_3}}{(d-3) s_{12}^2 s_{13}} , \nonumber \\
        -P_2^{\mu_1 \mu_2 \mu_3} &= -\frac{(d-4) T_1^{\mu_1 \mu_2 \mu_3}}{(d-3) s_{12}^2 s_{23}}-\frac{d s_{13} T_2^{\mu_1 \mu_2 \mu_3}}{(d-3) s_{12} s_{23}^3}+\frac{(d-4) T_3^{\mu_1 \mu_2 \mu_3}}{(d-3) s_{12} s_{13} s_{23}}\nonumber \\
        &-\frac{(d-2) T_4^{\mu_1 \mu_2 \mu_3}}{(d-3) s_{12} s_{23}^2} , \nonumber \\
        -P_3^{\mu_1 \mu_2 \mu_3} &= \frac{(d-4) s_{23} T_1^{\mu_1 \mu_2 \mu_3}}{(d-3) s_{12}^2 s_{13}^2}+\frac{(d-4) T_2^{\mu_1 \mu_2 \mu_3}}{(d-3) s_{12} s_{13} s_{23}}-\frac{d s_{23} T_3^{\mu_1 \mu_2 \mu_3}}{(d-3) s_{12} s_{13}^3}\nonumber \\
        &+\frac{(d-2) T_4^{\mu_1 \mu_2 \mu_3}}{(d-3) s_{12} s_{13}^2} , \nonumber \\
        -P_4^{\mu_1 \mu_2 \mu_3} &= -\frac{(d-2) T_1^{\mu_1 \mu_2 \mu_3}}{(d-3) s_{12}^2 s_{13}}-\frac{(d-2) T_2^{\mu_1 \mu_2 \mu_3}}{(d-3) s_{12} s_{23}^2}+\frac{(d-2) T_3^{\mu_1 \mu_2 \mu_3}}{(d-3) s_{12} s_{13}^2}\nonumber \\
        &-\frac{d T_4^{\mu_1 \mu_2 \mu_3}}{(d-3) s_{12} s_{13} s_{23}} \ , \label{eq:projectors_ggHg}
\end{align}
where $d=4-2 \epsilon$. In particular, for projecting the amplitude $A_{\vcr \vcr}$ onto the $T_i$, we choose the reference momenta to be $q_{1(2)}=p_{2(1)}$ and $q_3=p_1$. 
The tensor-structures  \cref{eq:tensor_structures_ggHg} agree with the ones given in \cite{Gehrmann:2011aa} under re-labeling, e.g. the same ordering for solving the overdetermined system in the determination of the independent $T_i$ was used.

\subsection{Functional relations} \label{subsec:func_rel}
The amplitude $A_{\vcr \vcr}$ is the leading-order amplitude for the
partonic process $g g \to Hg$ with the electroweak loop. Therefore, it is free
of explicit poles in the dimensional regulator and only has implicit singularities in the
IR-singular configurations. However, due to the IBP-reductions and the
projections, the obtained result has spurious explicit poles multiplying
different Laurent coefficients of the canonical MIs $f_{i}$. The semi-analytic integration we employ (see \cref{subsec:series_exp_method}) does not give a result in terms of special functions and, therefore, functional relations are not explicit, e.g. we do not have manifest cancellations of spurious poles in the dimensional regulator. In the following, we outline how the explicit functional relations can be obtained by exploiting the fact that our set of MIs are canonical.  This can directly be
used to impose cancellation of spurious poles and it simplifies the amplitude considerably. \\
For this we consider the explicit (spurious) pole of order $j$ of the scalar form-factor:
\begin{align}
    A^{(-j)} = \frac{1}{\epsilon^j} \sum_{k} \sum_{i_k} \alpha_{j,k,i_k}(\vec{x}) f_{i_k}^{(k)}  
\end{align}
where $\alpha_{j,k,i_k}(\vec{x})$ is an algebraic function of the external scales and $f_{i_k}^{(k)} $ is the $k$-th Laurent coefficient of the $i_k$-th canonical integral. In order for this pole to vanish, we first notice that the Laurent coefficients $f_{i_k}^{(k)}$ are transcendental functions of weight $k$. This directly implies that there is a set of $\mathbb{Q}$-linear independent algebraic functions $\alpha_{j,m,k}(\vec{x})$, such that 
\begin{align}
    A^{(-j)} = \frac{1}{\epsilon^j} \sum_k \sum_{m} \alpha_{j,m,k}(\vec{x}) \left(\sum_{i_m} q_{i_m} f_{i_m}^{(k)} \right) \label{eq:explicit_pole}
\end{align}
where $q_{i_m}\in \mathbb{Q}$, and 
\begin{align}
     A^{(-j)} = 0 \;\;\; \Rightarrow \;\;\;  F_{m} = \left(\sum_{i_m} q_{i_m} f_{i_m}^{(k)} \right) = 0 , \ \forall m . \label{eq:funct_equations} 
\end{align}
$F_{m}$ is a functional relation between Laurent coefficients of weight $k$. \\
In order to obtain the functional relations $F_m$ between the weight $k$-coefficients of the canonical MIs we proceed as follows. For each spurious pole of the amplitude $A^{(-j)}$ we obtain the analytic coefficients 
$\vec{\tilde{\alpha}}_{j,k}$ in front of all weight $k$ integrals such that 
\begin{align}
    A^{(-j)} = \frac{1}{\epsilon^j}  \sum_{k} \left( \vec{\tilde{\alpha}}_{j,k} \cdot \vec{f}^{(k)} \right) ,
\end{align}
where $\cdot$ is the usual scalar product. We then build a matrix $A_{\text{num.}}$ where each row is $\vec{\tilde{\alpha}}_{j,k}$ evaluated at $(s/m_V^2,t/m_V^2,m_H^2/m_V^2) = \vec{\pi}= (\pi_1,\pi_2,\pi_3)$ where the $\pi_i$ are different and sufficiently large prime numbers. The sampling is performed for more (distinct) prime-tuples $\vec{\pi}$ than there are weight $k$ integrals in  $A^{(-j)}$. The row-reduction $A_{\text{num.}}^{\text{row-red.}}$ of $A_{\text{num.}}$ will have $n$ non-zero rows with elements $q_{i_m}\in \mathbb{Q}$, each corresponding to one of the functional relations \cref{eq:funct_equations}:
\begin{align}
  F_{m} = \left(\sum_{i_m} q_{i_m} f_{i_m}^{(k)} \right) = 0 \;\;\; \Leftrightarrow \;\;\; \exists \, l : \sum_{i_l} \left(A_{\text{num.}}^{\text{row-red.}} \right)_{l,i_{l}}  f^{(k)}_{i_l} = F_m = 0 .
\end{align}
We use this sampling to reveal 444 functional identities between Laurent coefficients of scalar integrals up to weight 3, effectively by-passing the analytic decomposition \cref{eq:explicit_pole}. This
approach is very efficient, since row-reductions of large numeric matrices is not a bottleneck. We then impose these relations to make the vanishing of the spurious poles manifest, which, as a by-product, simplifies the finite remainder considerably.

\subsection{Validation of the amplitude} \label{subsec:AVV_checks}

In order to validate our result for the two-loop amplitude for $g g \to Hg$ we performed a variety of checks.\\
First, we verified
that our amplitude reproduces numerically the $g g \to Hg$-HEFT one in
the limit $m_V\to \infty$ as predicted by the infinite boson mass approximation
\cite{Anastasiou:2008tj}. \\ 
Furthermore, we validated
the interference of our amplitude against the interference obtained from the helicity
amplitudes provided in \cite{Bonetti:2020hqh}. This could only be done in the
Euclidean regime, since in order to obtain the full set of helicity amplitudes
from \cite{Bonetti:2020hqh} one needs to relabel their results, which amounts to a kinematic
crossing into a regime for which \cite{Bonetti:2020hqh} does not provide the analytic
continuation. \\
Lastly, in \cite{Becchetti:2020wof} we used our amplitude to compute the light quark-contribution of the NLO mixed QCD-electroweak contribution to the gluon fusion Higgs production cross section. In this computation,
we perform our phase-space integration with two
different local subtraction schemes (see \cite{Becchetti:2020wof} for more details), both of which are fully automatized,
generic, and independent implementations. As a consequence errors occurring in the amplitude expression would have been seen as missed numerical
cancellations in the IR, which do not occur.
This is a powerful additional check since it happens independently at runtime
and for physical kinematics.
\section{Computation of the Master Integrals}
\label{sec:MIs}
In this section we discuss the computation of the MIs that appear in the amplitude $\mathcal{A}$ \cref{eq:amp}, which has been performed by means of the differential equation approach \cite{Kotikov:1990kg,Remiddi:1997ny, Gehrmann:2002zr}  together with the generalized series method \cite{Francesco:2019yqt} in order to obtain numerical results for MIs in the physical region. For the computation we heavily rely on the publicly available IBP-reduction programs Kira~\cite{Maierhoefer:2017hyi,Maierhofer:2018gpa,Klappert:2020nbg}, Fire~\cite{Smirnov:2008iw,Smirnov:2013dia,Smirnov:2014hma,Smirnov:2019qkx}, and LiteRed \cite{Lee:2012cn,Lee:2013mka}.
\par
If one allows for the permutations:
\begin{align}
	\begin{array}{llll}
	\text{Map I:}&p_1\to p_1 & p_2\to -p_3 & p_3\to -p_2, \\
	\text{Map II:}&p_1\to p_2 & p_2\to p_1 & p_3\to p_3 ,\\
	\text{Map III:}&p_1\to -p_3 & p_2\to p_2 & p_3\to -p_1, \\
	\end{array}
\end{align}
all the scalar integrals appearing in the $A_i$ in \cref{eq:vector_vector_piece} can be considered as (sub-)topologies of the propagator sets shown in \cref{tab:propagator_sets}.
\begin{table}[htbp!]
	\centering
	\caption{Propagator sets for the scalar topologies in $gg \to Hg$. External momenta are denoted by $p_i$ and loop-momenta by $k_i$, where $p_i^2=0$, $ (p_1+p_2-p_3)^2=m_H^2$, and $m_V$ denotes the gauge boson masses $m_Z$ or $m_W$. \label{tab:propagator_sets} }
	\begin{tabular}{l|l}
		$P$ & $NP$ \\ \hline
		$D_{1,P}=k_1^2 $                                & $D_{1,NP}= k_1^2 $\\
		$D_{2,P}=\left(k_1-k_2\right){}^2 $             & $D_{2,NP}=\left(k_1-k_2\right){}^2 $\\
		$D_{3,P}=\left(k_1+p_1\right){}^2 $             & $D_{3,NP}= \left(k_1+p_1\right){}^2 $\\
		$D_{4,P}=\left(k_1+p_1+p_2\right){}^2 $         & $D_{4,NP}= \left(k_1+p_1+p_2\right){}^2 $\\
		$D_{5,P}=\left(-k_1-p_3\right){}^2 $            & $D_{5,NP}= \left(k_1-k_2-p_3\right){}^2 $\\
		$D_{6,P}=\left(k_2+p_1+p_2\right){}^2-m_V^2 $   & $D_{6,NP}= \left(k_2+p_1+p_2\right){}^2-m_V^2 $\\
		$D_{7,P}=\left(-k_2-p_3\right){}^2-m_V^2 $      & $D_{7,NP}= \left(-k_2-p_3\right){}^2-m_V^2 $\\
		$D_{8,P}=\left(k_1-k_2-p_3\right){}^2 $         & $D_{8,NP}= \left(-k_1-p_3\right){}^2 $\\
		$D_{9,P}=\left(k_2+p_2\right){}^2 $             & $D_{9,NP}= \left(k_2+p_2\right){}^2 $
\end{tabular}
\end{table}
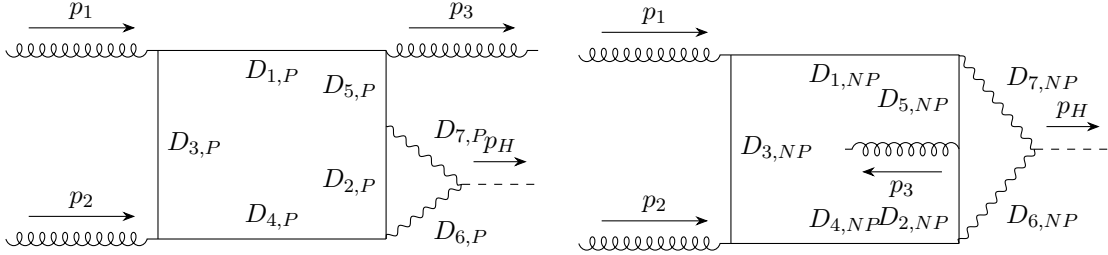
\begin{figure}[h!]
	\begin{center}
		\begin{subfigure}{0.49\textwidth}
			\centering
			\scalebox{1.}{\begin{tikzpicture}[scale=0.9, every node/.style={scale=0.9}]
				\begin{feynman}
				\vertex (in1) ;
				\vertex[below=2.5cm of in1] (in2) ;
				\vertex[right=2.0cm of in1] (a1);
				\vertex[below=2.5cm of a1] (a2);
				\vertex[right=3.0cm of a2] (a3);
				\vertex[above=1.5cm of a3] (a4);
				\vertex[above=1.0cm of a4] (a5);
				\vertex[right =1.cm of a3] (fake1);
				\vertex[above =0.725cm of fake1] (b1);
			    \vertex[right = 1.cm of b1] (outh);
			    \vertex[right = 2.cm of a5] (outg);
			    
				\diagram*{
				    (in1) -- [gluon, momentum = $p_1$] (a1),
				    (in2) -- [gluon, momentum = $p_2$] (a2),
					(a3) -- [plain, edge label =$D_{2,P}$](a4),
					(a2)-- [plain, edge label =$D_{4,P}$](a3),
					(a1)-- [plain, edge label =$D_{3,P}$](a2),
					(a4) -- [plain, edge label =$D_{5,P}$](a5),
					(a5) -- [plain, edge label =$D_{1,P}$](a1),
					(a5) -- [gluon, momentum = $p_3$](outg),
					(b1) -- [photon, edge label = $D_{6,P}$](a3),
					(a4) -- [photon, edge label = $D_{7,P}$](b1),
					(b1) -- [scalar, momentum = $p_H$](outh),
				};
				\end{feynman}
				\end{tikzpicture}}	
		\end{subfigure}
		\begin{subfigure}{0.49\textwidth}
			\centering
			\scalebox{1.}{\begin{tikzpicture}[scale=0.9, every node/.style={scale=0.9}]
				\begin{feynman}
				\vertex (in1) ;
				\vertex[below=2.5cm of in1] (in2) ;
				\vertex[right=2.0cm of in1] (a1);
				\vertex[below=2.5cm of a1] (a2);
				\vertex[right=3.0cm of a2] (a3);
				\vertex[above=1.25cm of a3] (a4);
				\vertex[above=1.25cm of a4] (a5);
				\vertex[right =1.cm of a3] (fake1);
				\vertex[above =1.25cm of fake1] (b1);
			    \vertex[right = 1.cm of b1] (outh);
			    \vertex[left = 1.5cm of a4] (outg);
			    
				\diagram*{
				    (in1) -- [gluon, momentum = $p_1$] (a1),
				    (in2) -- [gluon, momentum = $p_2$] (a2),
					(a3) -- [plain, edge label =$D_{2,NP}$,near start](a4),
					(a2)-- [plain, edge label =$D_{4,NP}$](a3),
					(a1)-- [plain, edge label =$D_{3,NP}$](a2),
					(a4) -- [plain, edge label =$D_{5,NP}$](a5),
					(a5) -- [plain, edge label =$D_{1,NP}$](a1),
					(a4) -- [gluon, momentum = $p_3$,near end](outg),
					(b1) -- [photon, edge label = $D_{6,NP}$](a3),
					(a5) -- [photon, edge label = $D_{7,NP}$](b1),
					(b1) -- [scalar, momentum = $p_H$](outh),
				};
				\end{feynman}
				\end{tikzpicture}}	
		\end{subfigure}
	\end{center}
	\caption{The scalar top-topologies associated to the propagator sets in \cref{tab:propagator_sets}. Continuous lines denote massless propagators, wavy lines denote massive propagators. All external particles are on-shell. \label{fig:toptopos}}
\end{figure}

The associated graphs, corresponding to the propagators $D_{1,(P,NP)},\dots D_{7,(P,NP)}$ are depicted in \cref{fig:toptopos}. Here, continuous lines denote massless propagators and wavy lines correspond to the massive gauge-boson propagators.\par 
The two-loop scalar integrals can be written in dimensional regularization as
\begin{equation}
\label{integral}
   \mathcal{J}^{(P,NP)}_{a_1,\cdots,a_9} =  \int \mathcal{D}^d k_1 \mathcal{D}^d k_2 \dfrac{D_{8,(P,NP)}^{a_8}D_{9,(P,NP)}^{a_9}}{D_{1,(P,NP)}^{a_1}\cdots D_{7,(P,NP)}^{a_7}},
\end{equation}
where $a_i \geq 0$ are positive integers, $d = 4 - 2\epsilon$ and the integration measure is defined as
\begin{equation}
    \mathcal{D}^d k_i = \dfrac{\dd^d k_i}{i \pi^{\frac{d}{2}}}e^{\epsilon \gamma_E} \left(\frac{m^2_V}{\mu^2}\right)^{\epsilon}.
\end{equation}

The number of master integrals for the planar topology (P) is 48, while for the non-planar topology (NP) we found 61 master integrals. The MIs relevant for this process have been analytically computed in \cite{Becchetti:2018xsk} and \cite{Bonetti:2020hqh}, respectively for the planar and non-planar topologies. For the purpose of this project we performed an independent numerical computation based on the generalized series expansion method described in \cite{Francesco:2019yqt}. The MIs basis, the differential equation matrices and the boundary values are given in the ancillary files, along with the scalar form-factors and the relevant tensor structures. The numerical evaluation is performed with a private implementation of the method \cite{Francesco:2019yqt}, however it is possible to use the material given in the ancillary files also within the software \textsc{DiffExp} \cite{Hidding:2020ytt}.

\subsection{Differential equations approach}

The MIs satisfy a system of linear first order partial differential equations with respect to the kinematic invariants \cite{Remiddi:1997ny,Gehrmann:1999as}. In order to solve the system efficiently we adopted the canonical basis approach \cite{Henn:2013pwa,Kotikov:1990kg}. A set of MIs $\vec{f}(\vec{x},\epsilon)$ is said to be in canonical form if it satisfies a system of differential equations of the kind:
\begin{equation}
    \label{eq:diffcan}
    \dd \vec{f}(\vec{x},\epsilon) = \epsilon \sum_i A_i(\vec{x}) \vec{f}(\vec{x},\epsilon) \dd x_i , \,\,\, A_i(\vec{x}) \coloneqq \partial_{x_i} \tilde{A} (\vec{x}),
\end{equation}
where $\epsilon$ is the dimensional regularization parameter and $\dd$ is the total differential with respect to the kinematic invariants $\vec{x}$. The matrix $\Tilde{A}(\vec{x})$ is a $\mathbb{Q}$-linear combination of logarithms,
\begin{equation}
    \label{eq:logmat}
    \Tilde{A}(\vec{x}) = \sum_{i} c_i \log (\alpha_i(\vec{x})),
\end{equation}
where $c_i$ are matrices of rational numbers and $\alpha_i(\vec{x})$ are algebraic functions of the kinematic invariants which we refer to as \emph{letters}. The set of letters is usually called the \emph{alphabet}. For the planar (non-planar) topology we find 25 (62) independent letters, such that
\begin{align}
\sum_{i}^{25, (62)} c_i \dd \log (\alpha_i(\vec{x})) = 0 \quad \Rightarrow \quad c_i=0 \ \forall i ,
\end{align}
which involve the square roots 
\begin{align}
    r_1&=\sqrt{-m_H^2} \sqrt{4 m_V^2-m_H^2}, \nonumber \\ 
    r_2&=\sqrt{s-m_H^2} \sqrt{-m_H^2+4 m_V^2+s},\nonumber \\
    r_3&=\sqrt{-t} \sqrt{4 m_V^2 (s+t) \left(m_H^2-s\right)-t m_H^4}, \nonumber\\
    r_4&=\sqrt{-m_H^2+s+t} \sqrt{4 m_V^2 \left(m_H^2-s\right) \left(m_H^2-t\right)+m_H^4  \left(-m_H^2+s+t\right)},
    \label{eq:square_roots}
\end{align}
where $s=(p_1+p_2)^2$ and $t=(p_1-p_3)^2$. The alphabet for the planar and non-planar topologies is provided as part of the ancillary material.

In order to find a canonical basis several approaches have been proposed~\cite{Henn:2014qga,Argeri:2014qva,Lee:2014ioa,Lee:2020zfb,Gituliar:2017vzm,Prausa:2017ltv,Dlapa:2020cwj}. We choose the approach discussed in \cite{Gehrmann:2014bfa,Becchetti:2017abb}.
The system of differential equations \cref{eq:diffcan} admits the solution in terms of Chen iterated integrals \cite{Chen:1977},
\begin{equation}
    \label{eq:chenint}
    \vec{f}(\vec{x},\epsilon) = \mathbb{P}\exp\left(\epsilon \int_{\gamma} \dd \Tilde{A}(\vec{x})\right) \vec{f}(\vec{x}_0,\epsilon),
\end{equation}
where $\mathbb{P}$ is the path-ordering operator, $\gamma$ represents a path in kinematic space and $\vec{f}(\vec{x}_0,\epsilon)$ is the vector of boundary conditions. In the context of high-energy physics, the solution \cref{eq:chenint} is written as a series expansion with respect to the dimensional regularization parameter around the point $\epsilon = 0$,
\begin{equation}
    \label{eq:solexp}
    \vec{f}(\vec{x},\epsilon) = \sum_{k = 0}^{\infty} \epsilon^k \vec{f}^{(k)}(\vec{x})
\end{equation}
The coefficients of the expansion can be explicitly written, for example, parametrizing the integration path $\gamma$ with a parameter $t \in \left[0,1\right]$:
\begin{equation}
    \label{eq:solexp2}
    \vec{f}(\vec{x},\epsilon) = \vec{f}^{(0)}(\vec{x}_0) + \sum_{k = 1}^{\infty}\epsilon^k\sum_{j = 1}^{k}\int_0^{1}\dd t_1 A(t_1)\int_0^{t_1}\dd t_2 A(t_2)\cdots\int_0^{t_{j-1}} \dd t_j A(t_j)\vec{f}^{(k-j)}(\vec{x}_0),
\end{equation}
where the matrices $A(t)$ denote the pull-back onto the interval $[0,t]$ along the path $\gamma$:
\begin{equation}
    A(t)\dd t := \gamma^* (\dd\Tilde{A}(\vec{x}))(t).
\end{equation}

The alphabet of the system of differential equations determines the functional space in which the solution to \cref{eq:diffcan} is represented. Specifically, if the $\alpha_i(\vec{x})$ are rational in the kinematic invariants, it is possible to write \cref{eq:solexp2} order-by-order in $\epsilon$ in terms of MPLs~\cite{Goncharov:1998kja,Goncharov:2001iea},
\begin{equation}
    \label{eq:gonchpollog}
    G(w_1,\cdots,w_n;z) = \int_0^z \dfrac{dt}{t-w_1}G(w_2,\cdots,w_n;z)\,,
\end{equation}
with
\begin{equation}
G(;z) = 1\,, \qquad G(\vec{0};z) \equiv \dfrac{\log^n z}{n!}\,.
\end{equation}

For algebraic $\alpha_i(\vec{x})$ in \cref{eq:logmat}, it is not always possible to write the solution of \cref{eq:diffcan} in terms of MPLs~\cite{Brown:2020rda}. However, in certain cases one can obtain a representation in terms of MPLs by employing computational techniques involving the \emph{symbol} associated to the solution itself~\cite{Goncharov:2010jf,Duhr:2011zq,Bonciani:2016qxi}. Such is the case of both the planar and the non-planar topologies of the two-loop $gg \to Hg$ amplitude~\cite{Becchetti:2018xsk,Bonetti:2020hqh}, for which a representation in terms of MPLs can be achieved, despite the fact that
the system of differential equations depends on the set of square roots \cref{eq:square_roots}.

\subsection{Series expansion method}
\label{subsec:series_exp_method}
In our computation of the mixed QCD-EW cross section \cite{Becchetti:2020wof}, we did not aim at a fully analytic two-loop $gg \to Hg$ amplitude as in \cite{Bonetti:2020hqh}.
Instead we exploited the method of generalized power series \cite{Francesco:2019yqt} to evaluate the MIs numerically  in the physical phase-space regions. In the following paragraph, we want to review the main points of this semi-analytical integration. \par 
The method can be summarised as follows: given the knowledge of the solution of the system \cref{eq:diffcan} in a given point $\vec{x}_0$ (either analytically or numerically with high-precision), it is possible to evaluate the solution in a point $\vec{x}_a$ numerically, by patching together local solutions in terms of generalized power series. The advantages of this strategy are several. First, it allows to obtain high-precision numerical results for a system of MIs regardless of the actual space of functions in which the analytic solution may be expressed. This aspect is particularly relevant when dealing with Feynman integrals that admit a solution in terms of elliptic integrals \cite{Bonciani:2016qxi,Bonciani:2019jyb,Frellesvig:2019byn}, for which the numerical evaluation of analytic expression has received, recently, increasing attention \cite{Duhr:2019rrs,Abreu:2019fgk,Walden:2020odh}. Secondly, even if the solution can be expressed in terms of MPLs it can be more convenient, especially for phenomenological applications \cite{Becchetti:2020wof,Abreu:2020jxa}, to exploit this generalized power series method. In particular when the system of differential equations involves square roots of the kinematic invariants, the numerical evaluation of an analytic solution expressed in terms of MPLs can be less efficient and it involves complicated analytic continuation for the special functions.\par
We assume that the solution to the system \cref{eq:diffcan} is known in a point $\vec{x}_0$ and we want to evaluate it at a point $\vec{x}_a$. The system \cref{eq:diffcan} can be written with respect to some variable $t$ which parametrize the path $\gamma(t)$ that connects the points $\vec{x}_0$ and $\vec{x}_a$:
\begin{equation}
\gamma(t) \, : \, t \mapsto \vec{x} (t), \,\,\, t \in \left[0,1\right], \,\,\, \gamma(0) = \vec{x}_0 \, , \, \gamma(1) = \vec{x}_a.
\end{equation}
As already shown in \cref{eq:solexp2}, the solution can be written as
\begin{align} 
& \vec{f}(t, \epsilon) = \sum_{k = 0}^{\infty} \epsilon^k \vec{f}^{(k)} (t), \\ \label{eq:expansion}
& \vec{f}^{(k)}(t) = \sum_{j = 1}^{k}\int_0^{1}\dd t_1 A(t_1)\int_0^{t_1}\dd t_2 A(t_2)\cdots\int_0^{t_{j-1}}\dd t_j A(t_j)\vec{f}^{(k-j)}(\vec{x}_0)+\vec{f}^{(k)}(\vec{x}_0).  
\end{align}
The first step is to split the path which connects the points $\vec{x}_0$ and $\vec{x}_a$ into segments $S_i \coloneqq \left[t_i - r_i, t_i + r_i\right)$. Then, inside each segment $S_i$, we denote by $\vec{f}^{(k)}_i (t)$ the local solution obtained as a truncated power series expansion around $t_i$ with radius of convergence $r_i$. The global solution on the path $\gamma(t)$ can then be approximated as:
\begin{align} \label{eq:solutionexpanded}
\vec{f}(t, \epsilon) = \sum_{k=0}^{\infty} \epsilon^k \sum_{i = 0}^{N - 1}\rho_i (t) \vec{f}^{(k)}_i (t), \;\;\; \rho(t)  = 
\begin{cases}
1, & t \in \left[t_i -r_i, t_i + r_i\right) \\
0, & t \notin \left[t_i -r_i, t_i + r_i\right)
\end{cases},
\end{align}
where $N$ is the total number of segments.\par
We can construct the segments $S_i$ from the knowledge of the singular points of the differential equations. In the general case, we can have both real and complex-valued singular points. Let us denote the real singular points of the system of differential equations as $R \coloneqq \left\{ \tau_i  \,\,\, \vert \,\,\, i = 1,\cdots,N_r \right\}$, and the complex-valued ones as $C \coloneqq \left\{\lambda_i^{re} + i \lambda_i^{im} \,\,\, \vert \,\,\, i = 1,\cdots, N_c\right\}$. Starting from $C$ we can construct a set of regular points $C_r \coloneqq \cup_{i=1}^{N_c} \left\{\lambda_i^{re} \pm \lambda_i^{im} \right\}$, therefore we can choose the expansion points $t_i$, which define the segments $S_i$, to belong to the set $R \cup C_r$, and the radius of convergence $r_i$ can be defined as the distance of $t_i$ to the closest element $t_p$, with $p \neq i$.\par
As a second step we need to know the local solution inside each segment $S_i$. This can be done by expanding the system of differential equations around the point $t_i$:
\begin{equation} \label{eq:sysexp}
A(t) = \sum_{l=0}^{\infty} A_l \left(t-t_i\right)^{w_l}, \,\,\, w_l \in \mathbb{Q},    
\end{equation}
where $A_l$ are constant matrices. By substituting \cref{eq:sysexp} into $\vec{f}^{(k)}_i (t)$ we obtain:
\begin{align} \label{eq:local1}
\vec{f}_i^{(k)}(t) =& \sum_{j=1}^k \sum_{l_1=0}^{\infty}\cdots\sum_{l_j=0}^{\infty} A_{l_1} \cdots A_{l_j} \int_0^t \dd t_1\left(t_1-t_i\right)^{w_{l_1}}\cdots \int_0^{t_{j-1}} \dd t_j \left(t_j - t_i\right)^{w_{l_j}}\vec{f}_i^{(k-j)}(\vec{x}_0) \nonumber \\
&+\vec{f}_i^{(k)}(\vec{x}_0).
\end{align}
As a consequence of working with a system of differential equations in canonical form, the integrals that appear in \cref{eq:local1} are of the form:
\begin{equation}
\int_0^{t_0} \dd t \left(t - t_i\right)^w \log\left(t - t_i\right)^m, \,\,\, w\in \mathbb{Q}, \,\, m\in \mathbb{N},    
\end{equation}
this implies that \cref{eq:local1} can be written as
\begin{equation}
\vec{f}_i^{(k)}(t) = \sum_{l_1 = 0}^{\infty}\sum_{l_2 = 0}^{N_{i,k}} c_k^{(i,l_1,l_2)} \left(t - t_i\right)^{\frac{l_1}{2}} \log(t-t_i)^{l_2}, \label{eq:expanded_laurent_coefficient}
\end{equation}
where the rational exponent $\frac{l_1}{2}$ is a consequence of the presence of the square roots, \eg \cref{eq:square_roots}, in the system of differential equations. The matrices $c_k^{(i,l_1,l_2)}$ depend on the boundary conditions and the constant matrices $A_l$ in \cref{eq:sysexp}. Finally, using the previous result for $\vec{f}_i^{(k)}(t)$, it is possible to numerically evaluate the solution in the point $\vec{x}_a$ by employing the expression \cref{eq:solutionexpanded}.\par
We conclude this brief review of the method by emphasizing that the endpoint of the integration path is considered a fixed numerical value so that the result is not considered a function of the endpoint itself, as for example in fully analytic approaches.

\par
\subsection{Analytic continuation}
\label{subsec:anal_cont}
Let us now turn to the analytic continuation, which in this approach simply means
a singularity of the matrix $A(t)$ exists for $t \in [0,1]$ such that the contour will
cross a branch-cut. In \cref{eq:expanded_laurent_coefficient} it is obvious that
we will need to analytically continue only roots (typically square roots) and
logarithms.
For physical thresholds $s_{phys.}$, which are linear in one
kinematic variable $x_k$, Feynman prescription dictates to assign a small
positive imaginary part $i \eta$, such that
\begin{align}
    x_k(t) = a  + b t \mapsto a  + b t + i \eta &  & \Rightarrow &  &
    x_k(t) \mapsto x_k(t + i \text{sign}(b) \eta) . \label{eq:ana_cont_t}
\end{align}
This directly determines the imaginary part for the logarithms of the series
around the threshold. More subtle are the singularities of the differential equations
which are multivariate polynomials in the kinematic invariants. For these
singularities, an analytic continuation can be very non-trivial often. However, if
these singularities are spurious the analytic continuation for intermediate
steps does not matter. We say, a singularity is spurious if all the numeric coefficient
$c_k^{(i,l_1,l_2)}$ in \cref{eq:expanded_laurent_coefficient} vanish identically; the
Feynman integral does not dependent on these logarithms at all. This can be
turned into an efficient check. We assume that all multivariate polynomial
singularities are spurious. Then we check explicitly, for each integration, that indeed all numeric coefficients
vanish whenever a contour crosses such singularities. \\
The analytic continuation for ``physical square roots'' (linear in one kinematic
invariant) is completely analogous. However, a sufficient condition for a
square-root to be spurious does not only involve the vanishing of the numeric
coefficients $c$ in \cref{eq:expanded_laurent_coefficient}. This is due to the
algebraic transformation we use to obtain a canonical basis $\vec{f}$ from a
basis of scalar integrals $\vec{\mathcal{J}}$
\begin{align}
    \vec{f} = T(\vec{x},\epsilon) \vec{\mathcal{J}} .
\end{align}
For a square root to be spurious all coefficients in front of that root
have to vanish for all $\mathcal{J}_i \in \vec{\mathcal{J}}$. So whenever a contour crosses an
assumed spurious branch-cut of a root, one has to invert back to the
scalar integrals $\vec{\mathcal{J}}$ and verify that indeed all coefficients in front of the
root vanish for all scalar Feynman integrals. For logarithms, this is not
necessary since they will never be used in the transformations to obtain a
canonical basis. \\

\subsection{Boundary conditions}
In order to evaluate the semi-analytic integration we obtain the initial boundary value in the large boson mass limit $m_V^2 \gg |m_H^2|, |s|, |t|$ when approached along a straight line from the un-physical region $m_H^2, s , t < 0$. In this limit, only a subset of factorized one-loop and sunrise-type integrals contribute. In particular, we compute the integrals:
\begin{align}
&\mathcal{J}^{(NP)}_{000210200}     , && 
\mathcal{J}^{(NP)}_{000211000}      , &&
\mathcal{J}^{(NP)}_{000220100}      , &&
\mathcal{J}^{(NP)}_{002012000}, \nonumber \\
&\mathcal{J}^{(NP)}_{002021000}     , &&
\mathcal{J}^{(NP)}_{020100200}      , &&
\mathcal{J}^{(NP)}_{020200100}      , &&
\mathcal{J}^{(NP)}_{021000200}, \nonumber \\
&\mathcal{J}^{(NP)}_{022000100}     , &&
\mathcal{J}^{(NP)}_{200100200}      , &&
\mathcal{J}^{(NP)}_{200101200}       &&
\end{align}
and respectively:
\begin{align}
&\mathcal{J}^{(P)}_{000210200} , &&
\mathcal{J}^{(P)}_{000220100} , &&
\mathcal{J}^{(P)}_{002010200} , &&
\mathcal{J}^{(P)}_{020012000}, \nonumber \\
&\mathcal{J}^{(P)}_{020021000} , &&
\mathcal{J}^{(P)}_{021000200} , &&
\mathcal{J}^{(P)}_{022000100} , &&
\mathcal{J}^{(P)}_{101110200}, \nonumber \\
&\mathcal{J}^{(P)}_{200100200} , &&
\mathcal{J}^{(P)}_{210000100} , &&
\mathcal{J}^{(P)}_{210002000} , &&
\mathcal{J}^{(P)}_{220001000} 
\end{align}
exactly and augment them with the large mass expansion of: 
\begin{align}
\mathcal{J}^{(NP)}_{101112000} , && 
\mathcal{J}^{(NP)}_{101121000} , && 
\mathcal{J}^{(NP)}_{111100200} , && 
\mathcal{J}^{(NP)}_{121100100} 
\end{align}
and: 
\begin{align}
\mathcal{J}^{(P)}_{111012000} , &&
\mathcal{J}^{(P)}_{111100200} , &&
\mathcal{J}^{(P)}_{121011000} , &&
\mathcal{J}^{(P)}_{121100100} 
\end{align}
respectively. The large mass expansion is performed by means of an expansion by regions with help of the code \textsc{ASY} \cite{Jantzen:2012mw}. The boundary conditions are included in the ancillary file.

\section{Conclusions}
\label{sec:NLO}
The two-loop amplitude described in this work has been used for our computation of the light-quark contribution to the NLO-mixed QCD-electroweak contribution to the gluon fusion Higgs production cross section \cite{Becchetti:2020wof}. This means, in particular, that our result is aimed at sufficiently fast,  high-precision, numerical evaluations in the complete physical phase-space and it motivates our choice of the semi-analytic integration discussed in \cref{subsec:series_exp_method}. In order to perform the integration we use a private implementation of the algorithm \cite{Francesco:2019yqt}. In particular, to facilitate the cross section computation, we have pre-computed a grid of values obtained by an analogous $g g \to Hg $-HEFT computation. This means that our starting grid is non-uniform and much more dense near the IR-singular regions.\par
The evaluation of the $\sim 300$ canonical integrals (including crossing), takes on average $\sim 1~ \text{min}$ on a single CPU thread. For phase-space points very close to the IR-singular configuration or deep in the UV the evaluation time increases considerably to $\sim 5-15~\text{min}$. In these regions, the scale hierarchies become extreme and higher depth expansions are needed in order to reach a fixed precision of better than $16$-digits. \\
In order to perform the cross section computation \cite{Becchetti:2020wof} we integrated the amplitude into \textsc{MadGraph5\_aMC@NLO} program \cite{Alwall:2014hca},
allowing for the generation of a standalone library for
the evaluation of all matrix elements entering the computation, by means of a dedicated plugin\footnote{available under \\
\url{https://bitbucket.org/aschweitzer/mg5_higgs_ew_plugin/}\\
or\\ \url{http://madgraph.physics.illinois.edu/Downloads/PLUGIN/higgsew.tar.gz}} detailed in \cref{app:plugin}. \\
We furthermore provide ancillary files for the differential equations, the boundary conditions and the amplitude, which allows a standalone evaluation of the amplitude within the publicly available code \textsc{DiffExp} \cite{Hidding:2020ytt}.

\section{Acknowledgements}
The authors are grateful to Roberto Bonciani and Vittorio del Duca for useful inputs and discussions throughout the project. We also thank Valerio Casconi for the valuable work in the early stage of the computation.
M.B. also acknowledges the financial support from the European Union Horizon 2020 research and innovation programme: High precision multi-jet dynamics at the LHC (grant agreement no. 772009).

\appendix

\section{Example usage of the dedicated \textsc{MadGraph5\_aMC@NLO}-plugin}
\label{app:plugin}
The $A_{gg \to Hg}^{\text{QCD-EW}}$-amplitude contributes to a
pure
NLO-QCD correction,  even though the LO is at two-loops. In particular, this
means it is suited for a fully automatized treatment in publicly available NLO matrix element generators. For the cross section computation \cite{Becchetti:2020wof}, we implemented our amplitude as a plugin into the {\normalfont \scshape\small MadGraph5\_aMC@NLO}
programme~\cite{Alwall:2014hca}, which we will abbreviate as \mgamc. In the
\mgamc-framework Feynman rules are implemented in the Universal FeynRules Output
({\normalfont \scshape\small UFO})~\cite{Degrande:2011ua} model. The \UFO~is a
representation of a vertex $v$ in the following universal form 
\begin{align}
    v(\text{fields},\text{momenta}) = \sum_{i} \sum_j \left(\text{color fact.}\right)_i \times \left(\text{Lorentz struct.}\right)_j \times \left(\text{coupling const.}\right)_{ij} .
\end{align}
In the usual pipeline this representation is derived directly from the
Lagrangian with the FeynRules-package. In \mgamc~this representation is used as
follows. One defines the process and specifies the coupling order. Then all
valid amplitudes at the specified coupling order are build, the Feynman rules
get inserted and the resulting expression auto-generates FORTRAN code for the
numerical evaluation with the help of \ALOHA~\cite{deAquino:2011ub}. Each
matrix-element becomes a standalone FORTRAN-library that is used for all
successive computations. \\
The \UFO-model is our entry point into this pipeline. All color-decomposed,
tensor-projected amplitudes define an effective vertex where the coupling
constants are the scalar form-factors. The only difference w.r.t. to the
tree-level \UFO-representation is that our coupling constant gets dynamically
updated at each phase-space point. We define for each mixed QCD-EW $A_{gg\to
H}^{\text{QCD-EW}}$, $A_{gg\to H}^{\text{QCD-EW}}$, $A_{gg\to H+g}^{\text{QCD-EW}}$, and for each
HEFT-amplitude $A_{gg\to H}^{\text{HEFT}}$, $A_{gg\to H}^{\text{HEFT}}$, $A_{gg\to
H+g}^{\text{HEFT}}$ an effective vertex. The virtual mixed QCD-EW amplitude are taken from \cite{Bonetti:2017ovy,Bonetti:2018ukf}. 
For performance reasons we fix the Higgs and
the electroweak-boson masses such that the virtual amplitudes are just a
number\footnote{We validate that the input parameters correspond to our fixed
values. \Eg when our implementation is run with masses different to these
values, it will abort and inform the user. However, we provide these coupling
calls as templates such that arbitrary mass-dependent computations can be
performed with very minor modifications of the code.}. The HEFT-amplitude
$A_{gg\to H+g}^{\text{HEFT}}$ is a pure-Lorentz structure, and only the mixed
amplitude $A_{gg\to H+g}^{\text{QCD-EW}}$ is a costly coupling computation at each
phase-space point. 

After following the installation instructions\footnote{available under \\
\url{https://bitbucket.org/aschweitzer/mg5_higgs_ew_plugin/}\\
or\\ \url{http://madgraph.physics.illinois.edu/Downloads/PLUGIN/higgsew.tar.gz}}
one can for example use our plugin with 
\begin{lstlisting}
    #
    # start mg5_aMC with our plugin
    #
    python2 ./mg5_aMC --mode=higgsew
    #
    # information on coupling parameters gets printed
    #
    GGHG.Interface: PLUGIN INFORMATION:
    GGHG.Interface: ----------------------------------
    GGHG.Interface: HEFT QCD-Background:
    GGHG.Interface: ----------------------------------
    GGHG.Interface: For LO: 
    GGHG.Interface:  generate g g > H   GGHEFT^2==2  QCD^2==4
    GGHG.Interface:  generate g g > H g GGGHEFT^2==2 QCD^2==6
    GGHG.Interface: For NLO-virtuals: 
    GGHG.Interface:  generate g g > H   GGHEFT^2==2  QCD^2==6
    GGHG.Interface: ----------------------------------
    GGHG.Interface: Massive Mixed EW:
    GGHG.Interface: ----------------------------------
    GGHG.Interface: For LO: 
    GGHG.Interface:  generate g g > H   GGHEFT^2==1  GGHEW^2==1  QCD^2==4
    GGHG.Interface:  generate g g > H g GGGHEFT^2==1 GGGHEW^2==1 QCD^2==6
    GGHG.Interface: For NLO-virtuals: 
    GGHG.Interface:  generate g g > H   GGHEFT^2==1  GGHEW^2==1  QCD^2==6
    GGHG.Interface: ----------------------------------
    GGHG.Interface: OUTPUT
    GGHG.Interface: ----------------------------------
    GGHG.Interface: use: output standalone_ggHg OUTPUTDIR
\end{lstlisting}

The generation statements are directly related to the \UFO-representation. One
defines a process and a coupling order. For example, for the HEFT-amplitudes we
see the definition 
\begin{lstlisting}
    GGHG.Interface: For LO: 
    GGHG.Interface:  generate g g > H    GGHEFT^2==2  QCD^2==4
\end{lstlisting}
The \textit{generate g g $>$ H} command defines the process, and
\textsc{GGHEFT\^{}2==2~~QCD\^{}2==4} specifies that we are working at
$C_{\text{Wilson}}^2$ and $g_s^4$. If we want one higher order in QCD, \eg
the virtuals, we need to increase to \textsc{QCD\^{}2==6} but retain the count
of the Wilson coefficient. If we interfere against non-HEFT amplitudes, the
squared Wilson coefficient count is decreased to \textsc{GGHEFT\^{}2==1}, \eg
only one of the interfered amplitudes involves a HEFT-coupling. One is now in
the default \mgamc~command line environment and can generate all processes. For
example, typing
\begin{lstlisting}
    MG5_aMC> generate g g > H g  GGGHEFT^2==1 GGGHEW^2==1 QCD^2==6
    #
    # printed info on amplitude
    #
    INFO: Process has 3 diagrams
\end{lstlisting}
will generate the amplitudes for the interferences of the real radiation
diagrams, where \prompt{} is just the prompt of the interface. The three
diagrams are the HEFT-amplitude and the mixed QCD-EW amplitudes for the two
different masses $m_{W/Z}$. In order to compile the process into a standalone
library one needs to specify an output directory 
\begin{lstlisting}
    MG5_aMC>output standalone_ggHg example_gghg_standalone
\end{lstlisting}
This output directory ``\textit{example\_gghg\_standalone}'' includes, among other things, a standalone library for the
numerical evaluation of the matrix elements. This means, one can perform the
cross section computation with any public or private code by linking against
this library. Since our scalar integrals are evaluated in \textsc{Mathematica},
which provides a poor interface with low-level languages, we perform a
offline-parallelisation over the Monte-Carlo grids of the cross section
computation. Sample codes which can be used to facilitate such a offline parallelization are provided with the plugin.

\newpage
\bibliographystyle{JHEP}
\bibliography{biblio}

\end{document}